\documentclass[namedreferences]{kluwer}
\usepackage{graphicx}
\usepackage{epsfig}
\usepackage{url}
\usepackage{amsbsy}
\usepackage{latexsym}
\usepackage[mathscr]{euscript}
\begin{document}
\begin{article}
\begin{opening}
\title{Multi-wavelength analysis of high energy electrons
in solar flares: a case study of August 20, 2002 flare}
\author{Jana \surname{Ka\v{s}parov\'{a}}}
\author{Marian \surname{Karlick\'{y}}}
\institute{Astronomical Institute, Academy of Sciences of the Czech Republic,\\
251 65 Ond\v{r}ejov, Czech Republic}
\author{Eduard P. \surname{Kontar}}
\institute{Department of Physics and Astronomy, University of Glasgow, \\Glasgow G12 8QQ, UK}
\author{Richard A. \surname{Schwartz}}
\institute{Science Systems \& Applications, Inc. at NASA Goddard Space Flight Center, Greenbelt, MD 20771, USA}
\author{Brian R. \surname{Dennis}}
\institute{NASA Goddard Space Flight Center, Greenbelt, MD 20771, USA}
\runningauthor{J. Ka\v{s}parov\'{a} et al.}
\runningtitle{Multi-wavelength analysis of the August 20, 2002 flare}
\begin{abstract}
A multi-wavelength spatial and temporal analysis of solar high
energy electrons is conducted using the August 20, 2002 flare of an
unusually flat ($\gamma_1=1.8$) hard X-ray spectrum. The flare is
studied using RHESSI, H$\alpha$, radio, TRACE, and MDI
observations with advanced methods and techniques never previously
applied in the solar flare context. A new method to account for X-ray
Compton backscattering in the photosphere (photospheric albedo) has been used to 
deduce the
primary X-ray flare spectra. The mean electron flux distribution
has been analysed using both forward fitting and model independent
inversion methods of spectral analysis. We show that the
contribution of the photospheric albedo to the photon spectrum
modifies the calculated mean electron flux distribution, mainly at
energies below $\sim$100~keV.
The positions of the H$\alpha$ emission and hard X-ray sources with
respect to the current-free extrapolation of the MDI photospheric
magnetic field and the characteristics of the radio emission
provide evidence of the closed geometry of the magnetic field structure and
the flare process in  low altitude magnetic loops. In agreement with
the predictions of some solar flare models, the hard X-ray sources
are located on the external edges of the H$\alpha$ emission and
show chromospheric plasma heated by the non-thermal electrons.
The fast changes of H$\alpha$ intensities are located not only inside
the hard X-ray sources, as expected if they are the signatures of
the chromospheric response to the electron bombardment, but also
away from them.
\end{abstract}
\end{opening}
\section{Introduction}
Analyses of the impulsive phase of solar flares detected in various
electromagnetic wavelength ranges are important for understanding the
acceleration and propagation of particles. The accelerated electrons 
reveal their presence e.g. in hard X-ray and  radio emissions, and
cause rapid heating of the chromosphere and subsequent emission in
the hydrogen H$\alpha$ line.

One of the key questions is the form of the accelerated electron 
distributions. Their determination from the hard X-ray spectra depends
not only on the emission mechanism but also on the processes affecting the propagation of the
particles. As suggested by Brown, Emslie, and Kontar (2003), it is the mean electron flux
distribution $\bar{F}(E)$ depending only on the bremsstrahlung cross section
that should be used as a reference for both observational hard X-ray spectral analyses and theoretical
models of electron acceleration and propagation. The mean electron flux
distribution can be determined either by inverting the photon spectra (Johns and Lin, 1992;
Piana et al., 2003) or by a forward fitting method usually assuming a power-law form of $\bar{F}(E)$
(Holman et~al., 2003). Observed hard X-ray spectra also contain a contribution of photons
backscattered in the photosphere and this will modify the calculated $\bar{F}(E)$ (Bai and Ramaty, 1978;
Alexander and Brown, 2002).

Besides the X-ray emission, numerous types of radio emission in the metric and decimetric range
are usually observed during flares. They provide further information about the
acceleration and propagation of the electron beams and plasma parameters 
in the emission region~(Karlick\'{y}, 1997).

According to standard solar flare models (e.g. Dennis and Schwartz, 1989),
the accelerated electrons provide one of the mechanisms for the flare energy 
transport from the release site to
the lower atmospheric layers. As the accelerated electrons propagate along
the magnetic field lines toward the photosphere, they lose their kinetic
energy mainly via Coulomb collisions (Brown, 1971) in the lower corona and chromospheric layers.
The chromospheric response to the beam energy deposition determines 
the characteristics of optical and UV emission from the flare loops.
Numerical models of chromospheric response to pulsed electron beam heating
(e.g. Canfield and Gayley, 1987; Heinzel, 1991) predict the
time correlation of hard X-ray and H$\alpha$ emission 
recently analysed e.g. by Trottet et al. (2000) and Wang et al. (2000).
Their results show that hard X-rays and H$\alpha$ intensities in some flare
kernels exhibit time correlations in the time range from subseconds to~$\sim$10~s.

Some flare models (e.g. Heyvaerts, Priest, and Rust, 1977; Cargill and Priest, 1983) and
observations (Czaykowska et al., 1999) suggest that
H$\alpha$ kernels are located between the upflows of the beam heated plasma and the downflows
of the cool plasma in the loops disconnected from the reconnection site.
The leading edges of the H$\alpha$ emission close to the newly reconnected loops are supposed to be also heated
by the accelerated particles, which are detectable e.g. as the hard X-ray sources.
Comparison of the spatial distribution of Yohkoh hard
X-ray sources and H$\alpha$ flare kernels was done by  e.g. Asai
et al. (2002). They found that many H$\alpha$ kernels brighten
successively during the evolution of the flare ribbons, but only a
few radiation sources were seen in the hard X-ray images.
They suggest that this discrepancy may be the result of the low dynamic range capability
of the Yohkoh Hard X-ray Telescope.

In this paper, we analyse the impulsive phase of the August 20, 2002 solar
flare, which is characterised by a very flat photon spectrum at the
X-ray burst maximum (spectral index $\gamma_{1}=1.8$). Similar flat spectra have been
reported, e.g. by Nitta, Dennis, and Kiplinger (1990) and F\'arn\'ik, Hudson, and Watanabe (1997).
The event presented here is the first such flat spectrum 
spectrum observed with the high energy resolution of the Reuven Ramaty
High Energy Solar Spectroscopic Imager (RHESSI) (Lin et al., 2002).
We show the influence of photons backscattered in the photosphere on the determination of the mean electron
flux distribution and compare results obtained by inversion and forward fitting methods.
We also study the spatial correlation of H$\alpha$ emission and its time changes with hard X-ray sources.
In section 2 we describe the global behaviour of the event with
regard to hard X-ray and radio fluxes. Sections 3 and 4 present the
RHESSI, H$\alpha$ and magnetic field observations, methods, and results.
Finally, the results are discussed in section 5 and the conclusions are given in section 6.
\section{Global event description}
The analysed flare is well suited for a multi-wavelength study of the electron propagation
because its observations provide a set of simultaneous data of comparable time and spatial resolution
in the wavelengths directly related to the accelerated particles: RHESSI hard X-ray spectra and images, 
H$\alpha$ images (Kanzelh\"{o}he observatory),
and radio and microwave fluxes (Bern, Ond\v{r}ejov, and Z\"{u}rich facilities).

The flare was detected by GOES satellites on August 20, 2002.
It started at 08:25~UT and reached its maximum at 08:26~UT as an M3.4
flare. The 1B H$\alpha$ flare was
reported in  NOAA AR 0069 at S10W38, starting at 08:25~UT, peaking at
08:26~UT, and ending at 
08:37~UT\footnote{\url{http://www.sec.noaa.gov/weekly/pdf2002/prf1408.pdf}}.

\begin{figure}[p]
\centerline{\includegraphics[width=0.94\textwidth]{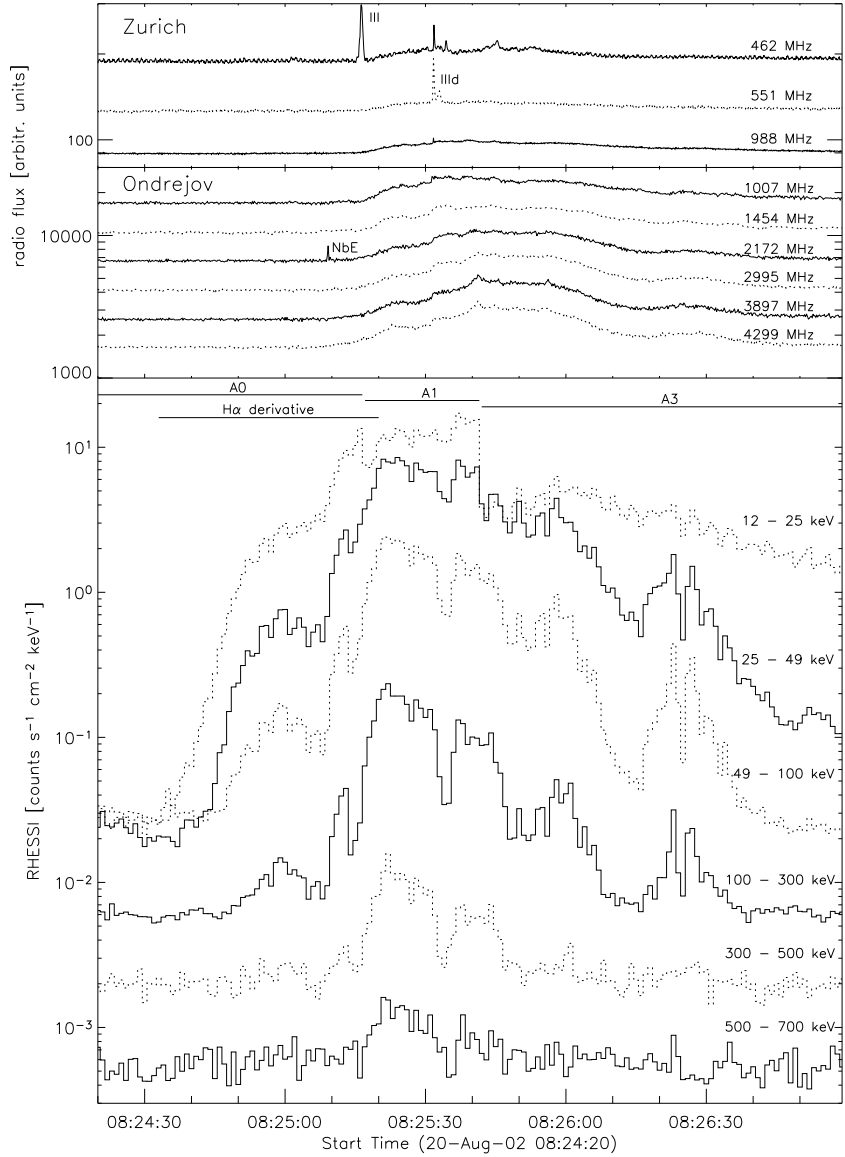}}
\caption{Time evolution of radio and hard X-ray fluxes
   during the August 20, 2002 flare from observations made with 
   Z\"{u}rich and Ond\v{r}ejov radiospectrometers and RHESSI.
   Narrowband emission (NbE), type III, and IIId
   bursts are annotated on the Ond\v{r}ejov and Z\"{u}rich data plots.
   For display purposes the RHESSI count flux at the energy band 
   300~-~500~keV was multiplied
   by a factor of 2, radio fluxes were shifted with a step of
   $\log\Delta=0.2$ from the fluxes at 4299~MHz (Ond\v{r}ejov) and 988~MHz (Z\"{u}rich).
   Attenuators states 
   A0, A1, and A3 and the time interval in which the derivative of H$\alpha$ intensity
   was analysed are indicated by the lines at the top of the RHESSI data plot.
   The abrupt changes in RHESSI flux in the 12~-~25~keV energy
   band at $\sim$~08:25:17~UT and $\sim$ 08:25:41~UT are caused by the
   attenuator changes (A0~$\rightarrow$~A1 and A1~$\rightarrow$~A3,
   respectively) at these times.}
\label{fig_hsi_radio}
\end{figure}
A global overview of the flare evolution in dm/m radio waves and
RHESSI X-rays is shown in Figure~\ref{fig_hsi_radio}. The RHESSI
flare was first detected at 08:24:32~UT with the increase in flux in the
12~-~25~keV energy band.  Then, with some time delays (at
$\sim$~08:24:45~UT), bursts in higher and higher energy bands followed.
During the starting phase of the X-ray emission at 08:24:30~-~08:25:08~UT,
no radio bursts were observed in the dm/m range.
The first weak and narrowband emission (NbE) in the dm-range was registered at 08:25:08~UT in the
2000~-~2200~MHz frequency range, followed by the type III burst below
500~MHz at 08:25:16~UT.

The most energetic part of this flare occurred during the time
interval 08:25:15~-~08:25:50~UT, when significant hard X-ray emission
was detected up to 7~MeV (see also data from the SONG instrument on board CORONAS-F; 
Myagkova, priv. comm.) and microwave emission up to 89.4~GHz.
The turn-over frequency between optically thick and thin part of the radio emission
was shifted to values between 19.6 and 50~GHz.
At that time the radio emission in the dm-range also reached the maximum and
a very fast drifting ($\sim$~-4000~MHz s$^{-1}$) and short-lasting
($\sim$~0.1~s) type IIId burst (relativistic type III burst - see
Poquerusse, 1994) was observed in the 400~-~1000~MHz range at
08:25:31.8~UT.  In the decimetric range (2000~-~4500~MHz) this phase
was characterised by broadband pulsations.
\section{RHESSI data analysis}\label{sec_rhessidata} RHESSI X-ray data
were analysed with the RHESSI Data Analysis Software
(Hurford et al., 2002; Schwartz et al., 2002). During the flare,
different attenuators were automatically placed in front of all detectors to absorb
soft X-rays and reduce pulse pile-up. The attenuator states A0, A1,
and A3 (no attenuator, thin attenuator, and both thick and thin
attenuator, respectively) are indicated on
Figure~\ref{fig_hsi_radio}. Corrections for the effects of the
different attenuator states are available for all times except when
the attenuators are moving, a period of $\sim$~1~s.
\subsection{Imaging} Image reconstruction was performed with both the
CLEAN and the PIXON algorithms using data from the front segments of the detectors. 
Detector 7 was used only for the
images in the energy bands above 7~keV due to its energy threshold of
about 7~keV.
Detector 1 and 2 mainly added noise to the images indicating 
that there was no significant source structure with an angular scale finer than $\sim$ 4 arcsecond.
Therefore, these detectors were excluded.

The CLEAN algorithm was used for obtaining morphology and time
evolution of X-ray sources, see sections~\ref{spatial} and \ref{tder},
Figures~\ref{fig_ha_hsi_mdi_low}, \ref{fig_ha_hsi_mdi_high}, and \ref{fig_haderiv_hsi}. Since
there was no source component as large as or larger than FWHM of the detector 8
(107~arcsec), the imaging fields of view in all CLEAN images were set smaller than this FWHM
and consequently detectors 8 and 9 were also not used in the CLEAN image reconstruction.

The PIXON algorithm (Puetter and Pi\~{n}a 1994; Metcalf et al., 1996) is
known to suppress spurious sources and to have high photometric
accuracy (Metcalf et al., 1996; Alexander and Metcalf, 1997).
We have used the PIXON imaging
technique for imaging spectroscopy at the burst maximum, see section~\ref{flatspectrum}.
\subsection{Spectroscopy}\label{spectroscopy}
We analysed data summed over the eight front segments. 
Detector 2 was excluded since its energy threshold is about 20~keV and
its energy resolution is about 10~keV. The data were corrected for pulse pile-up and
decimation.
Full 2D detector response matrices, each corresponding to the applied attenuator state,
were used to convert input photon fluxes to count rates.  The response matrix
accounts for the efficiency and resolution of the detectors, the absorption by the attenuators, 
grids, and all other material above the detectors, and all
other known instrumental effects (Smith et al., 2002).

The time bin for RHESSI spectra was 2~s
($\approx$ a half rotation), which ensures that the rapid modulation
produced by the grids does not have any distorting effect on the spectra.
The RHESSI spectra were analysed in the time interval 08:25:10 - 08:26:00~UT and were fitted from 8~keV
up to typically 400~keV, depending on the signal-to-noise ratio.

RHESSI data of this flare were contaminated by an electron 
precipitation\footnote{\url{hsi_20020820_071620_rate.png} and \url{hsi_20020820_071620_part_rate.png}
at \url{http://hessi.ssl.berkeley.edu/hessidata/metadata/2002/08/20/}}
starting at 08:16~UT and detected up to 300~keV. Therefore, a special care was taken to estimate
the background due to the electron precipitation in the 12~-~300~keV energy range while 
standard techniques were used at other energies. In the 12~-~40~keV energy range, the background was estimated
using count rates from the onboard charged particle detector whose lower threshold (50~keV) is generally 
triggered during electron precipitation events (Smith et al., 2002). Analysing several precipitation events
which occured on other orbits  two days before and after the flare, we have found a 
good correlation between the particle count rate and the front count rate in this energy range. Therefore,
count rates from the particle detector can serve as a proxy for the bremsstrahlung of the 
precipitating electrons as detected by germanium detectors in this energy band.

At higher energies the germanium detector count 
rates during the precipitation event are generally dominated by the
bremsstrahlung that originates from the electrons interacting in the Earth's atmosphere
and generally does not follow the time history of the particle count rate (Smith, priv. comm.).
Having taken this into account, the precipitation component in the 40~-~300~keV energy range 
was estimated with a cubic fit to the shoulders of the precipitation event detected 
before and after the flare. 

The uncertainty of the estimated 
precipitation component in the 12~-~300~keV energy range was estimated to be $\leq 50\%$.
Despite this high value, any residual precipitation component in the background 
subtracted count spectra in the 12~-~300~keV energy range
is $\leq 10\%$ at the edges of the analysed time interval and decreases to several percent 
with the increasing flare flux. Thus, the precipitation event does not affect our results in the
analysed time and energy range.
\subsection{Flat photon spectrum at the flare maximum}\label{flatspectrum}
In order to evaluate the power-law indices of the photon spectra
and to make comparisons with previous  works, we adopted a standard
photon spectrum model $I$ consisting of a double power-law
non-thermal component $I_\mathrm{nt}$ (e.g. Lin and Schwartz, 1987) and a
Maxwellian isothermal component $I_\mathrm{th}$: $I=I_\mathrm{th} + I_\mathrm{nt}$.
\begin{equation}\label{bpow}
I_\mathrm{nt}(\epsilon) = \left\{\begin{array}{ll}
            a\epsilon^{-\gamma_1} & \epsilon < \epsilon_\mathrm{b}\\
            a\epsilon_\mathrm{b}^{-\gamma_{1}+\gamma_{2}}\epsilon^{-\gamma_2} &
\epsilon\geq \epsilon_\mathrm{b}
            \end{array}\right.\,,
\end{equation}
where $\epsilon$, $\epsilon_\mathrm{b}$, $\gamma_1$, $\gamma_2$,
and $a$ are the photon energy, the break energy in keV, the
power-law index below and above $\epsilon_\mathrm{b}$, and $a$ is for normalisation, respectively.
We used the forward fitting method and fitted the background subtracted
X-ray spectra with the model spectrum $I$. The model spectrum was
convolved with the full 2D detector response matrix to compute the model count spectrum.
Parameters of the best fit model were found by
searching for a minimum $\chi^2_\mathrm{min}$ between the measured and the model  count
rate spectrum.

The lowest value of the power-law index $\gamma_1$ was found at the
burst maximum as seen in $> 25\,\mbox{keV}$ X-rays, at 08:25:22~UT, when it dropped to 1.8. Note that this
value lies outside of the range of common values of photon power-law
index, $\langle 2,7\rangle$, as reported e.g. by Dennis (1985).
The flatness of the spectrum does not result from the contamination by a generally flat
spectrum of the precipitation event because the residual precipitation component in the background-subtracted 
count  spectrum was estimated to be $\leq 1\%$ in the whole energy range 
at 08:25:22~UT.
Hence, we believe that the flat spectrum is of solar origin.

This conclusion is further supported by the results from the imaging spectroscopy using the
PIXON imaging method (Hurford et al., 2002).
RHESSI PIXON images are constructed by finding the simplest model which is consistent
with the counts observed during the time interval. The  model used included a source image and a
component that accounts for the counts unmodulated by the grids. This additional component 
prevents the unmodulated counts from being included in the source image. 
Thus, the PIXON image is not contaminated by this background and the PIXON image reconstruction 
provides an independent method of determining the spatially integrated flare spectrum.
We reconstructed the PIXON images in several energy bands for
4.141~s ($\sim$~1~spin period) centred at the burst maximum (08:25:22~UT)
and determined a PIXON photon spectrum from the total photon fluxes above
the 10\% contour level of the corresponding PIXON image.
Since the uncertainty in the spectra from the image reconstruction 
is $\sim$10\% (Hurford, priv. comm.), and the PIXON photon spectrum agrees 
within uncertainties of the same order with the flat photon spectrum up to 300~keV 
(i.e. the highest energy for which the sources were detectable), we may conclude that 
the background subtraction procedure, see section~\ref{spectroscopy}, resulted in a 
solar count spectrum not significantly distorted by the precipitation component.
\subsection{Correction for photospheric albedo and its implications on mean electron flux distribution}
Primary photons with energies larger than 15~keV emitted downwards
have a high probability of being reflected due to
Compton backscattering in the photosphere. These albedo photons
then modify both the spectrum and the intensity of the observed hard X-rays.
According to Bai and Ramaty (1978), the reflectivity of a
power-law photon spectrum from the isotropic source, which is
assumed here, is a function of the photon energy and significantly
depends on the photon spectral index $\gamma$. The reflectivity
peaks at $\sim 35\,\mathrm{keV}$ and increases as $\gamma$
decreases. The total photon spectrum, that is the primary one plus
the component composed of the reflected photons, is then
characterised by a lower $\gamma$ than that of the primary spectrum.
For a spectral index $\gamma \sim 1.8$, the reflected photons at
$35$~keV can be as high as 80\% of the primary flux. 
The amount of backscattered photons generally depends on the anisotropy of the primary hard X-ray source and
the pitch-angle distribution of the beam electrons. 
But, the dependency is rather weak due to strong angular smoothing effect of 
the bremsstrahlung cross section. A detailed study of beam electron distributions is out
of scope of this paper. However, the assumption of isotropy of the hard X-ray source provides 
a lower limit for the photospheric albedo correction.

For this flare we studied the influence of the photospheric
albedo correction on the mean electron flux distribution $\bar{F}(E)$.
We used two approaches. First, $\bar{F}(E)$ was approximated by a double power-law function
\begin{equation}\label{ffthin}
\bar{F}(E) =\left\{\begin{array}{ll}
      AE^{-\delta_1} & E_\mathrm{low} \leq E < E_\mathrm{break}\\
      AE_\mathrm{break}^{-\delta_{1}+\delta_{2}}E^{-\delta_2} & E_\mathrm{break}\leq E \leq E_\mathrm{high}\\
           \end{array}\right.\,,
\end{equation}
where $E$, $E_\mathrm{low}$, $E_\mathrm{break}$, $E_\mathrm{high}$, $\delta_1$, $\delta_2$,
and $A$ are the electron energy, the low-energy cutoff, the break energy, the high-energy cutoff in keV,
the power-law index below and above $E_\mathrm{break}$, and normalisation, respectively.
The mean electron flux distribution represents the density weighted electron flux distribution in the source,
and its determination is equivalent to the determination of the electron flux distribution assuming thin-target
bremsstrahlung $I_\mathrm{thin}(\epsilon)$ (Brown, Emslie and Kontar, 2003)
\begin{equation}\label{ithin}
I_\mathrm{thin}(\epsilon) = \frac{\bar{n}V}{4\pi R^2}\int\limits_\epsilon^{E_\mathrm{high}}
\bar{F}(E)Q(\epsilon,E)\mathrm{d}E\,,
\end{equation}
where $\bar{n}$ is the mean intensity in the source volume $V$, $R=1\,\mbox{AU}$ is the distance of the
emitting source, and $Q(\epsilon,E)$ is the isotropic
cross section for bremsstrahlung (Haug, 1997). Using the forward fitting method described in
section~\ref{flatspectrum}, the background subtracted photon spectra were fitted
with a Maxwellian isothermal component $I_\mathrm{th}$ plus
the thin-target bremsstrahlung $I_\mathrm{thin}$ of the electron flux model
$\bar{F}(E)$ and the best-fit values of thermal
plasma and electron flux parameters (the fit provides $\bar{n} V \bar{F}(E)$) were obtained.
$E_\mathrm{high}$ was kept fixed at 7~MeV since
setting the parameter free did not significantly improve the fits.

Second, the mean electron flux distribution function was derived
by a regularised inversion procedure (Piana et al., 2003; Kontar et
al., 2004, 2005a) using the photon spectra (determined with the
forward fitting method) as input. Electron energies were
determined in the energy range typically 10~-~400~keV, the upper
energy depending on the signal-to-noise ratio.

\subsubsection{Photospheric albedo}
Background subtracted photon spectra were corrected for
photospheric albedo using the method of an angle dependent Green's
function for X-ray Compton backscattering (Kontar et al.,
2005b), where the Green's function represents the probability
density of backscattering of a photon of initial energy
$\epsilon_0$ into the observer's direction with energy $\epsilon$.
This approach is independent of the spectral characteristics of
the primary photon spectra and allows any form of the primary photon spectra
to be deduced, not just a power-law spectrum
\begin{figure}[t]
\centerline{\includegraphics[width=0.8\textwidth]{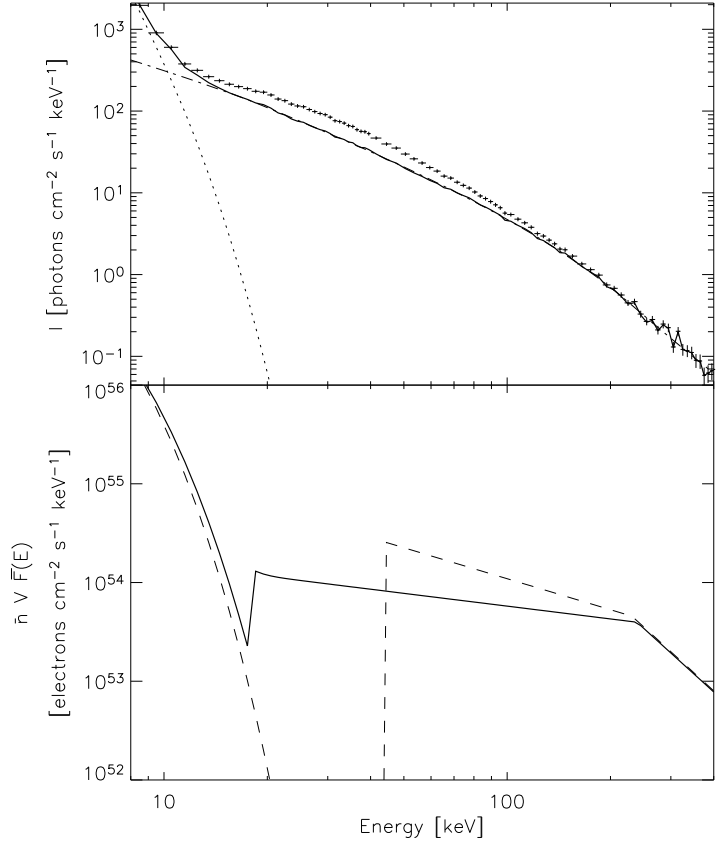}}
\caption{Influence of the photospheric albedo on the photon and mean electron flux spectrum at 08:25:22~UT
(intergration time 2s). Top panel shows the total (crosses) and primary photon spectrum 
(\vrule height 0.6ex depth -0.4ex width 1.5em). 
Horizontal sizes of crosses correspond to widths of energy bands, vertical sizes represent
$\pm 1\sigma$ statistical plus systematic uncertainties. Best fit to the primary photon flux is composed of
the isothermal $I_\mathrm{th}\,(\cdots)$ and thin-target $I_\mathrm{thin}\,(-\cdot-)$ component. 
Bottom panel compares forward fitted $\bar{n} V \bar{F}(E)$ derived from
the total (- - -) and primary (\vrule height 0.6ex depth -0.4ex width 1.5em) photon spectrum. 
Low-energy cutoffs and their standard
errors corresponding to the total and primary photon spectrum are
$E_\mathrm{low}^\mathrm{total}=44\pm 6\,\mathrm{keV}$ and $E_\mathrm{low}^\mathrm{pr}=20\pm\,^{10}_{20}\,\mathrm{keV}$,
respectively. See section~\ref{comp_ff_inv} for details on parameter error estimation.}
\label{elow_albedo}
\end{figure}
as was done e.g. by Bai and Ramaty (1978), and Alexander and Brown (2002). 
Green's function for a given heliocentric angle
(40 degrees for this flare) can be applied to the primary photon spectrum but 
as a computational convenience we modify the detector response matrix to account for the photospheric albedo.
In this way it can be used
straightforwardly in the forward fitting method 
(Kontar et al.\footnote{\url{http://www.astro.gla.ac.uk/users/eduard/rhessi/albedo/}})
because the result of the photon model
multiplied by this new modified detector matrix is exactly the same as the result from the
original detector matrix multiplied by the product of the albedo and the photon spectrum.

In addition to photospheric albedo, the RHESSI X-ray spectra generally contain a 
component from photons scattered in the Earth's atmosphere.
The geometry of RHESSI detectors is such that most of the photons scattered in Earth atmosphere must
first pass through the rear segments (see Figure 1 in Smith et al., 2002)., 
where the photons are effectively absorbed
(the stopping depth of Ge detectors for a photon of 30 keV is $\sim$ 1 mm).
Thus, only the photons coming from a forward semisphere can get
directly into the front segments. Since RHESSI is pointed towards
the Sun, the front segments are shielded by the rear ones from the Earth
scattered photons all the time when RHESSI can see the Sun apart from
being at the terminator. 
The studied flare was observed when RHESSI was far from the
terminator. Therefore, a correction for the Earth albedo photons for actual zenith angle was neglected
in the analysis.

The primary photon spectrum (derived under the assumption of its isotropy) 
together with the total photon spectrum at 08:25:22~UT (burst maximum) is shown 
in the top panel of Figure~\ref{elow_albedo}. The photon spectra differ mainly 
below 100~keV, the primary spectrum is slightly steeper, $\gamma_1^{\mathrm{pr}} = 1.9$, and
exhibits less flattening below 50~keV than the total photon spectrum. 
Such an albedo correction significantly
affects $\bar{n} V \bar{F}(E)$ mainly at electron energies below 100~keV. The bottom panel of
Figure~\ref{elow_albedo} shows that a forward fitted $\bar{n} V \bar{F}(E)$
consistent  with the total photon spectrum  has a high value of $E_\mathrm{low}^{\mathrm{total}}$, 
which would represent a gap
in energy between the thermal and electrons producing hard X-rays. However, removing 
the contribution of the photospheric albedo photons results in a forward fitted 
$\bar{n} V \bar{F}(E)$ with  $E_\mathrm{low}^{\mathrm{pr}}$ close to the photon energy where
the thermal part of the photon spectrum joins the non-thermal part. Consequently, 
no reliable dip in $\bar{n} V \bar{F}(E)$ is found.

\subsubsection{Comparison of forward fitting and inversion methods}\label{comp_ff_inv}
\begin{figure}[tbh]
\centerline{\includegraphics[width=0.7\textwidth]{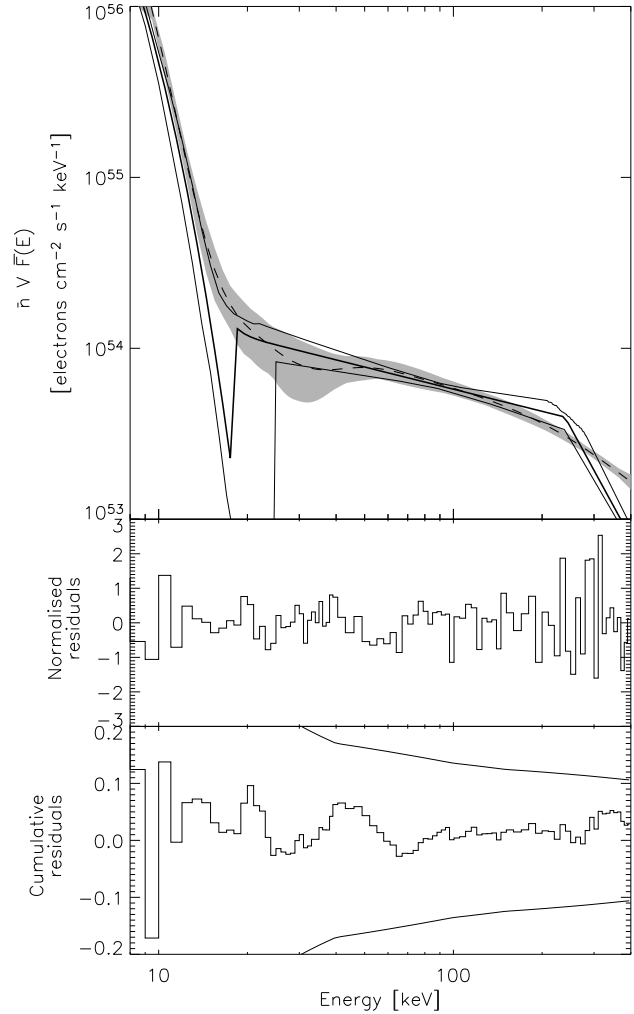}}
\caption{
Top panel shows comparison of $\bar{n} V \bar{F}(E)$ deduced from forward fitting 
(\vrule height 0.6ex depth -0.4ex width 1.5em) and inversion
({\bf - - -}) method for primary photon spectrum at 08:25:22~UT.
Shaded area represents $\pm$1-$\sigma$ limits on the regularised solution, thin full lines
correspond to 68\% confidence strip for the forward fit solution. Middle and bottom panels show normalised
and cumulative residuals of the primary and modelled photon spectrum for the forward fitting method.
Thin lines in the bottom panel represent $\pm$1-$\sigma$ limits for a random walk process.
}
\label{albedo_ff_inv}
\end{figure}

The mean electron flux distribution $\bar{n} V \bar{F}(E)$ corresponding to the primary photon spectrum
at 08:25:22~UT derived from both the forward fitting method and the regularised inversion procedure are shown
in the top panel of Figure~\ref{albedo_ff_inv}.
The ranges of parameter values were
determined as a change $|\chi^2-\chi^2_\mathrm{min}|\sim 8.2$ while keeping the individual parameter fixed,
where the level 8.2 encloses the region of joint confidence of 7 parameters with 68\% confidence 
(Lampton, Margon and Bowyer, 1976; Press et al., 1992). 
Uncertainties of the forward fitted $\bar{n} V \bar{F}(E)$ were determined by multiple generations
of the forward fit parameters lying within the same level of joint confidence.
$\pm$1-$\sigma$ limits of the regularised solution were found from multiple inversions
of the primary photon flux data perturbed within corresponding statistical and systematic uncertainties.
The inversion method was studied in detail  in Kontar et al.~(2004), where various modelled input spectra 
were analysed.
Middle and bottom panels of Figure~\ref{albedo_ff_inv}
represent normalised and cumulative residuals of the primary  $I^\mathrm{pr}$
and modelled photon spectrum $I=I_\mathrm{th}+I_\mathrm{thin}$. Normalised residuals are defined as
$r(\epsilon)=[I^\mathrm{pr}(\epsilon) - I(\epsilon)]/\sigma(\epsilon)$, where $\sigma(\epsilon)$ is the
uncertainty in the primary photon spectrum and includes statistical plus systematic uncertainties.
The error
of the albedo correction method is $\sim 6\%$, which is larger than $\sigma(\epsilon)$ but not included in it.
Cumulative residuals are defined as $S(\epsilon)=1/N\sum_i^N r_i$, where $N$ is the $N^\mathrm{th}$ energy bin
used in fitting,
and can be used to assess clustering of residuals in certain energy ranges (Piana et al., 2003).
Normalised residuals of the best fit are limited to a 3-$\sigma(\epsilon)$ interval. Cumulative residuals
show minor clustering in the photon energy range 20~-~80keV but they are well below $\pm$1-$\sigma$ limits of 
random walk
($\sigma=1/\sqrt N$), see full lines in bottom panel of Figure~\ref{albedo_ff_inv}. Both normalised and cumulative
residuals also indicate that assumed systematic uncertainties (5\% in each energy bin) are overestimated.
Adopting a lower value of 3\% results in residuals distributed as expected for 
Gaussian statistics with $\sim 1/3$ of them
above the 1-$\sigma$ level, but it does not modify  $\bar{n} V \bar{F}(E)$ significantly.

\begin{figure}[t]
\centerline{\includegraphics[width=0.94\textwidth]{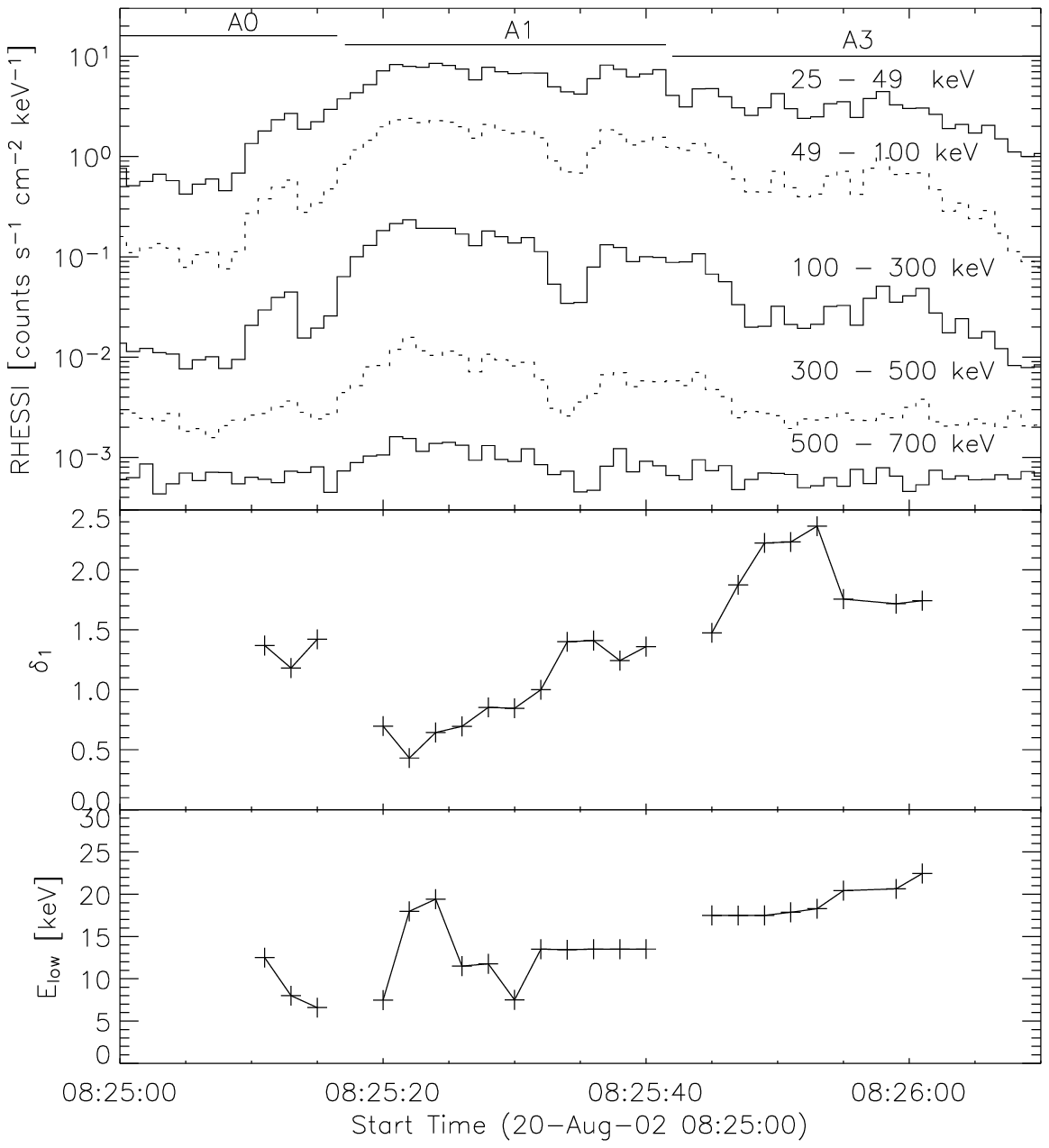}}
\caption{Time evolution of RHESSI X-ray count flux (1-s time resolution) in five energy channels,
       the power-law index $\delta_1$, and the low-energy cutoff $E_\mathrm{low}$ of the mean electron
    flux distribution $\bar{n} V \bar{F}(E)$. Attenuator states A0, A1, and A3 are indicated by lines at the top
       of RHESSI data plot. For display purposes the RHESSI count
    flux at the 300~-~500~keV band was scaled by a factor of 2.
}
\label{fthin_time}
\end{figure}
The best forward fit parameters for the primary spectrum are:
emission measure $EM=(1\pm\,^{2.0}_{0.4})\times 10^{49}\,\mathrm{cm}^{-3}$,
temperature $T = (1.3 \pm 0.2)\,\mathrm{keV}$,
$E_\mathrm{low}=(20\pm\,^{10}_{20})\,\mathrm{keV}$,
$\delta_1 = 0.4 \pm 0.2$, $\delta_2=3.2\pm\,^{0.3}_{0.2}$, $E_\mathrm{break} = (240\pm\,^{40}_{30})\,\mathrm{keV}$,
$\bar{n}V\bar{F}=(17\pm\,^{3.0}_{0.8})\times 10^{55}\,\mbox{electrons cm}^{-2}\,\mathrm{s}^{-1}$,
where $\bar{F}$ is $\bar{F}(E)$ integrated from $E_\mathrm{low}$ to $E_\mathrm{high}$.
Several parameters ($EM$, $E_\mathrm{low}$, $\bar{n}V\bar{F}$)
have significantly different, by a factor of 5, estimates of high and low $\sigma$.
This behaviour is mainly due to the low value of $\delta_1$ because a decrease in $E_\mathrm{low}$
and consequent increase in $\bar{n}V\bar{F}$ does not result in significant change in the modelled primary
photon spectrum.
The primary photon spectrum is therefore, within uncertainties, consistent also with a
mean electron flux distribution function without a low-energy cutoff as derived from the forward fitting method.
Large uncertainties in $EM$ are caused by the small number of data points in the thermal component of the
spectrum (see top panel of Figure~\ref{elow_albedo}).

The mean electron flux distribution deduced from the forward fitting 
agrees within uncertainties with the regularised solution up to $\sim$~200~keV. 
The discrepancy at higher energies is caused by different high-energy cutoffs in the
forward fitting and inversion methods. The regularised solution
was cut at 0.5~MeV whereas the double power-law $\bar{F}(E)$
extended to $E_\mathrm{high}=7\,\mathrm{MeV}$. 
The large uncertainties between
20 and 40 keV shown in Figure~\ref{albedo_ff_inv} result from the small number of electrons in
this energy range and the fact that the photons are primarily from higher
energy electrons. A similar argument can be used to explain the deviation
between the forward fit and regularised solutions above 100 keV. The high
value of  $E_\mathrm{high}$ used in the forward-fit solution introduces a contribution
to the photon flux above 100 keV from the higher energy electrons not
included in the regularised electron spectrum.

In the case of the total photon spectrum, i.e. without the albedo correction, the regularised solution
indicates a dip within 1-$\sigma$ limits in the energies below $\sim30$~keV
where the forward fitted solution resulted in a gap above the thermal component
in $\bar{n}V\bar{F}(E)$, see Figure~\ref{elow_albedo}. 
\subsubsection{Time evolution}
Figure~\ref{fthin_time} shows the time evolution of derived parameters $\delta_1$ and $E_\mathrm{low}$
of the mean electron flux distribution corresponding to the primary photon spectra.
The power-law index $\delta_1$ decreases to its minimum of $0.4 \pm 0.2$ at the burst maximum, 
08:25:22~UT, and its time
history generally shows a soft-hard-soft pattern between the photon flux and hardness of the spectra
(e.g. Fletcher and Hudson, 2002).
Low values of $\delta_1$ indicate a flat electron flux spectrum and this is confirmed by the regularised
inversion method.
The low-energy cutoff $E_\mathrm{low}$ lies within or close to the energies where the thermal component
dominates $\bar{n}V\bar{F}(E)$ distribution. Standard errors are of the order
of the $E_\mathrm{low}$ values. Therefore, no reliable dips or gaps in $\bar{n}V\bar{F}(E)$ throughout
the time interval could be established.
Our analysis shows that the low-energy cutoffs above the energies of the thermal component 
are likely to be the result of the contribution of photospheric 
albedo photons and do not exist in real mean electron spectra.

\section{H$\alpha$ data analysis and magnetic field topology}\label{sec_ha}
To study the influence of the accelerated electrons on the H$\alpha$ emission, we used
full-disk H$\alpha$ images covering the whole interval of the flare. The images were acquired using a filter
with 0.07~nm FWHM at Kanzelh\"{o}he Solar Observatory in Austria. The
time resolution of the data is 3~s and the spatial resolution is
2.2~arcsec/pixel. The position of the solar centre and the spatial
resolution were determined using the routines provided by A.~Veronig.
Details on the method can be found in Veronig et al. (2000).

\begin{figure}[p]
\centerline{\includegraphics[width=0.8\textwidth]{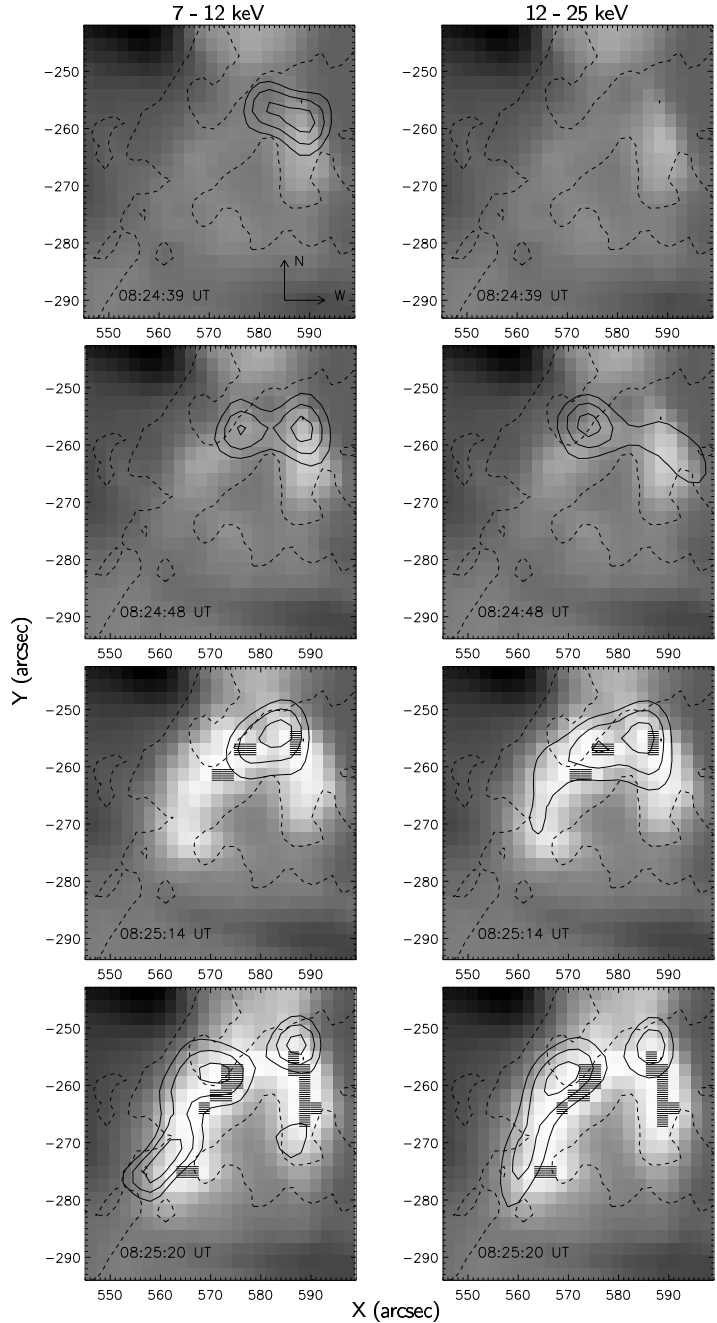}}
\caption{Time and spatial evolution of H$\alpha$ intensities (grey scale) and X-ray emission (full line contours)
in the 7~-~12~keV (first column) and 12~-~25~keV (second column) energy ranges.
Contours correspond to 50, 70, and 90 \% of maximum of each
RHESSI CLEAN image. There were no detectable X-ray sources in the 12~-~25~keV energy range
at 08:24:39~UT (see the image in the top right corner).
MDI magnetic neutral lines are plotted as  dashed lines.
The saturated H$\alpha$ pixels are displayed with horizontal lines.
}
\label{fig_ha_hsi_mdi_low}
\end{figure}

\begin{figure}[hbt]
\centerline{\includegraphics[width=0.8\textwidth]{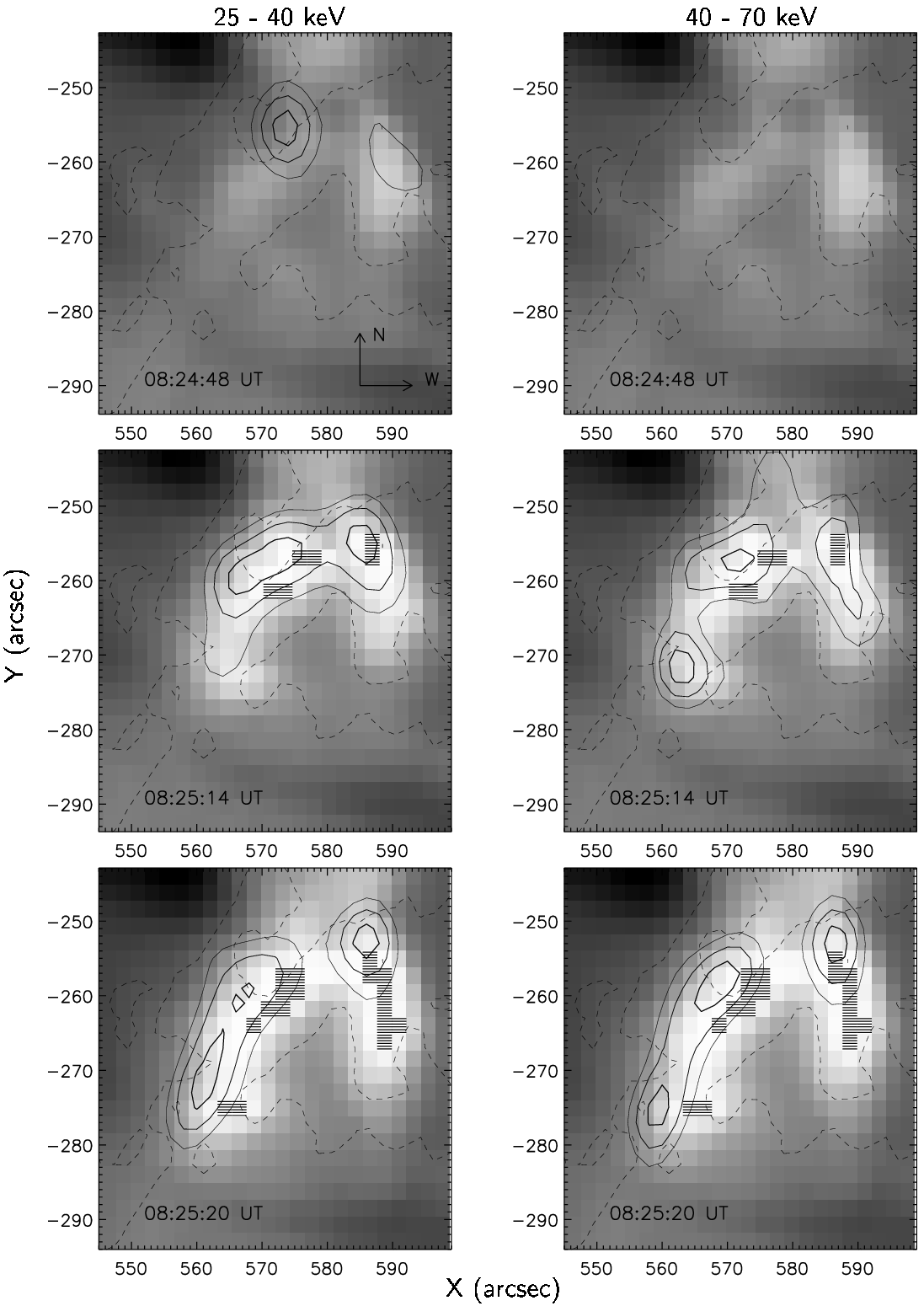}}
\caption{Time and spatial evolution of H$\alpha$ intensities (grey scale) and X-ray emission (full line contours)
in the 25~-~40~keV (first column) and 40~-~70~keV (second column) energy ranges.
Contours correspond to 50, 70, and 90 \% of maximum of each RHESSI CLEAN image. There were no
detectable X-ray sources in the
40~-~70~keV energy range at 08:24:48~UT (see the image in the top right corner).
MDI magnetic neutral lines are plotted as  dashed lines.
The saturated H$\alpha$ pixels are displayed with horizontal lines.
}
\label{fig_ha_hsi_mdi_high}
\end{figure}

The intensities from the flare site were scaled by the averaged quiet-Sun
area intensity I$_\mathrm{q}$ for every image separately. The brightest parts of the
flare area beginning at 08:25:20~UT are distorted by the saturation of
detected intensities.
The saturated pixels are displayed with horizontal lines in the H$\alpha$ images
(Figures~\ref{fig_ha_hsi_mdi_low}, \ref{fig_ha_hsi_mdi_high}, and \ref{fig_haderiv_hsi}).

The data set was further spatially coaligned using a technique developed
for TRACE and EIT data (Gallagher\footnote{\url{http://hesperia.gsfc.nasa.gov/~ptg/trace-align/index.html}}).
The cross-correlation of the
H$\alpha$ data set was done in
the area of the nearby sunspot with respect to the chosen H$\alpha$ image.
That sunspot area was not affected by the flare. Therefore, we assumed no time
change of the H$\alpha$ intensities during the analysed time interval in that area.
The positional uncertainty of the H$\alpha$ images was estimated to be of the order of 1 pixel (2.2~arcsec).
\subsection{Spatial evolution}\label{spatial}
The evolution of the hard X-ray sources overlayed with the H$\alpha$
emission and MDI magnetic neutral lines is displayed in Figures~\ref{fig_ha_hsi_mdi_low}
and \ref{fig_ha_hsi_mdi_high}.
The RHESSI flux was accumulated for 1~spin period (4.141~s).
The H$\alpha$ images correspond to the middle of the RHESSI imaging time range.
Due to the saturation of the H$\alpha$ intensities,
the spatial evolution in the flare area is shown to 08:25:20~UT only.

At the very beginning of the flare (at 08:24:39~UT, Figure~\ref{fig_ha_hsi_mdi_low}), a single
X-ray source is detected. Then it splits into two (at 08:24:48~UT, Figures~\ref{fig_ha_hsi_mdi_low}
and \ref{fig_ha_hsi_mdi_high}) and appears to be moving eastwards. Later at 08:25:14~UT,
the X-ray emission starts to shift southwards and the sources split even into three components detectable
in the energy range 40~-~70~keV. There is a difference in the positions of hard X-ray sources
at lower and higher energy ranges detected at the same time.
The X-ray sources at higher energies tend to be located more to the south than the lower energy X-ray sources
(compare the positions of the sources in different energy bands at 08:25:14~UT and 08:25:20~UT
in Figures~\ref{fig_ha_hsi_mdi_low} and \ref{fig_ha_hsi_mdi_high}).
Most of X-ray source maxima at energies above 12~keV
are not spatially coincident with the brightest parts of the H$\alpha$ emission.
Rather they tend to be located near the north and eastern edges of the bright H$\alpha$ kernels.
The H$\alpha$ kernels form a semi-circle whose inner boundary is located near the magnetic neutral line.
\subsection{Magnetic topology} 
\begin{figure}[ht]
\centerline{\includegraphics[width=0.65\textwidth]{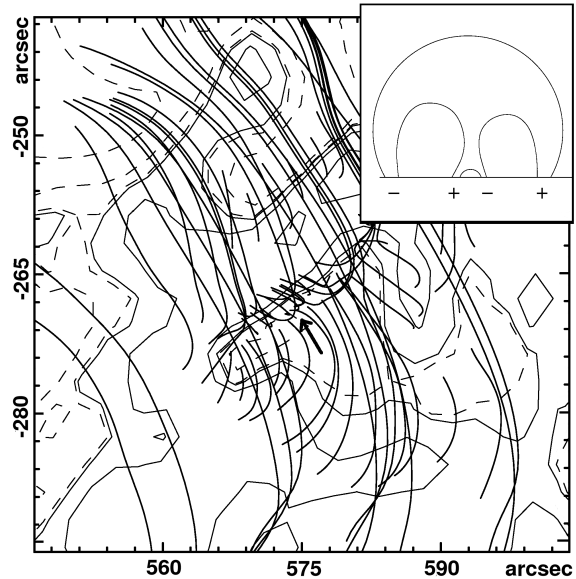}}
\caption{Magnetic field lines, extrapolated in the
      current-free approximation at 08:12:30~UT (full thick lines), revealing the
      quadrupolar topology near the flare site. The isolines of the north
      and south magnetic polarities ($\pm$~10, $\pm$~100, and $\pm$~400~G)
      are expressed by thin full and dashed lines, respectively.
      A cut through
     the region of the neutral magnetic line indicated by the arrow is
     shown in the sketch in the top right  corner (here sizes of loops are not to scale).}
\label{fig_mag}
\end{figure}

To understand better the magnetic
field topology near the flare site, the extrapolation of magnetic
field in the current-free approximation (D\'{e}moulin et al., 1997)
was made using the MDI magnetogram observed at 08:12:30~UT (see
Figure~\ref{fig_mag}).

The most interesting structure, indicated
by an arrow in Figure~\ref{fig_mag}, is located along the magnetic neutral line, which is
delineated by the above mentioned semi-circular H$\alpha$ ribbon.
Namely, this structure has a quadrupolar character in the sense of
papers by Uchida (1996) and Hirose and Uchida (2002), i.e.
low altitude loops in the close vicinity of the neutral line are between
magnetic field lines with the opposite orientation
and all is covered by extended overlying loops.

The preflare activity
in the 1550~\AA\ TRACE band  was observed  directly in the centre of
the quadrupolar structure (i.e. above the magnetic neutral line)
at 08:12:20~UT. However, no 3~-~6~keV X-ray emission was detected by
RHESSI from this flare site
at this time.

Comparing Figures~\ref{fig_ha_hsi_mdi_low}, \ref{fig_ha_hsi_mdi_high}  and \ref{fig_mag}, 
it can be seen that the
flare started at the north end of this quadrupolar structure. Then the
flare kernels appeared at the north-east boundary of the structure
and the flare ends at the southern extremity of the quadrupolar
structure. We also note that the H$\alpha$ ribbon and X-ray sources are shifted $\sim 4$~arcsec from the
quadrupolar configuration of the extrapolated magnetic field lines.
\subsection{Time derivative of H$\alpha$ intensities}\label{tder}
We assumed that the locations of the optical flare that are directly
affected by the  electron beams might be recognised as areas of
abrupt changes of the H$\alpha$ intensity, $I_{\mathrm{H}\alpha}$, on a time scale of the
order of seconds.
This assumption is supported e.g. by Trottet et al. (2000),
who found correlation between hard X-ray flux and H$\alpha$ intensities on time scales from
seconds to tens of seconds. Therefore, the time derivative of $I_{\mathrm{H}\alpha}$ may provide
a technique to find such areas.
\begin{figure}[t]
\centerline{\includegraphics[width=\textwidth]{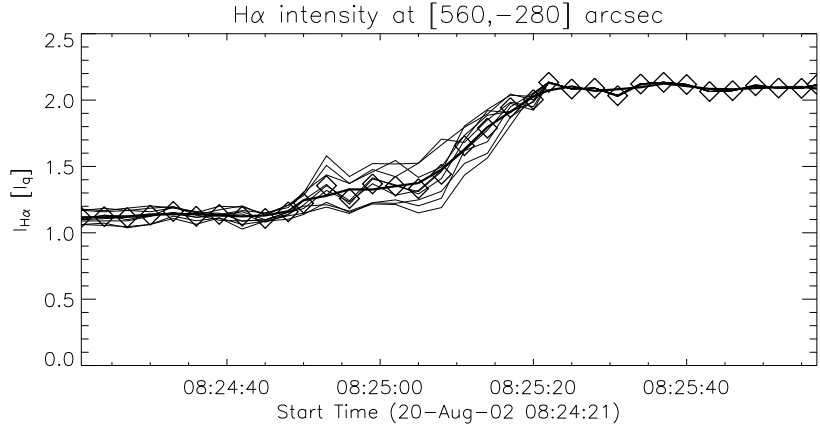}}	
\caption{Averaging and smoothing effects of $I_{\mathrm{H}\alpha}$ at the location [560,-280]~arcsec 
revealing the largest derivative of $I_{\mathrm{H}\alpha}$ (see Figure~\ref{fig_haderiv_hsi}). 
Thin lines show observed time histories at
3x3 pixel area, $\diamond$ correspond to the average over that area, and the thick line is the
resulting time history smoothed by 3-point box car. Intensities are scaled by the quiet Sun intensity 
$I_\mathrm{q}$.}
\label{fig_ha_time}
\end{figure}
\begin{figure}[t]
\centerline{\includegraphics[width=0.78\textwidth]{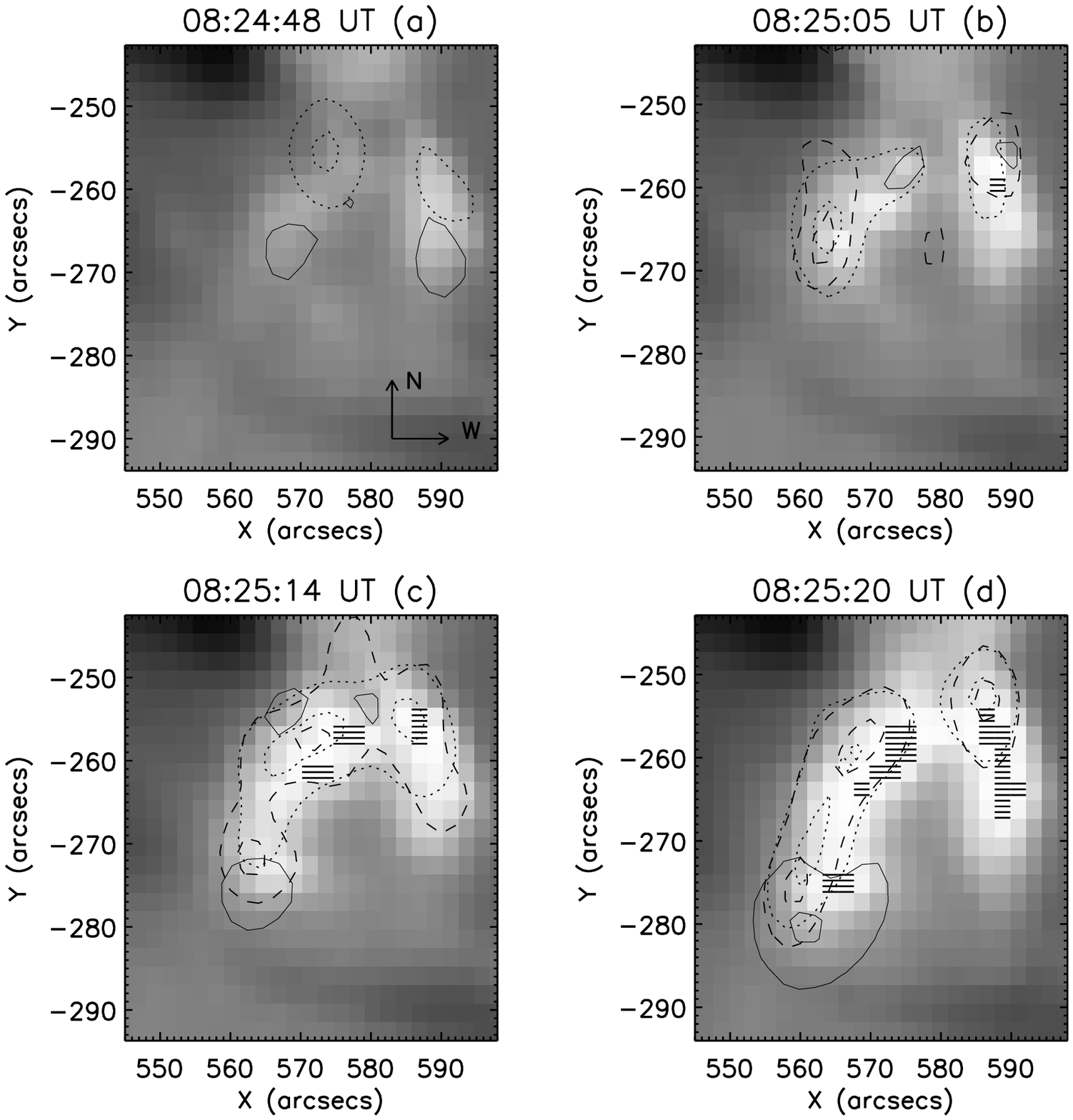}}
\caption{Locations of time derivative of H$\alpha$ intensities (full line contours) overlayed on
      H$\alpha$ images (grey scale). Contours correspond to
      $0.04\,\mathrm{I}_\mathrm{q}\,\mathrm{s}^{-1}$ and
      $0.09\,\mathrm{I}_\mathrm{q}\,\mathrm{s}^{-1}$ of
      $\mathrm{d}I_{\mathrm{H}\alpha}/\mathrm{d}t$. RHESSI CLEAN images
      at 25~-~40~keV and 40~-~70~keV are plotted as dotted and dashed
      contours at 50 and 90 \% of maximum of each image.
      The saturated H$\alpha$ pixels are displayed with horizontal lines.}
\label{fig_haderiv_hsi}
\end{figure}

In order to suppress the effects of seeing and/or incorrect positions,
H$\alpha$ intensities were averaged over a $3\times 3$ pixel area and the resulting
lightcurves were smoothed by a 3-point box car (9 s time resolution). Such smoothing was
necessary to distinguish between the real changes and those which occurred randomly
throughout the whole flare area. 
The effects of the averaging and smoothing are shown in Figure~\ref{fig_ha_time}.
Due to the saturation of H$\alpha$ data beginning at
08:25:20~UT, the time derivative of H$\alpha$ intensity, $\mathrm{d}I_{\mathrm{H}\alpha}/\mathrm{d}t$,
was studied in the time interval 08:24:33~-~08:25:20~UT (see also the label in Figure~\ref{fig_hsi_radio}).
The maximum of $\mathrm{d}I_{\mathrm{H}\alpha}/\mathrm{d}t$ was
$0.097\,\mathrm{I}_\mathrm{q}\,\mathrm{s}^{-1}$.
A threshold, $\mathrm{d}I_{\mathrm{H}\alpha}/\mathrm{d}t = 0.03\,\mathrm{I}_\mathrm{q}\,\mathrm{s}^{-1}$,
was estimated by fitting $\mathrm{d}I_{\mathrm{H}\alpha}/\mathrm{d}t$
distribution outside of the flare area with the Gauss distribution.
The threshold  corresponds to 10 standard deviations from the mean of the fit and
was adopted as the level of the reliable value of $\mathrm{d}I_{\mathrm{H}\alpha}/\mathrm{d}t$ not
affected by noise or seeing.

The locations of $\mathrm{d}I_{\mathrm{H}\alpha}/\mathrm{d}t > 0.03\,\mathrm{I}_\mathrm{q}\,\mathrm{s}^{-1}$
are displayed on Figure~\ref{fig_haderiv_hsi}
together with the RHESSI CLEAN images in the 25~-~40~keV and 40~-~70~keV energy bands.
X-ray sources in the 40~-~70~keV energy band are not shown in Figure~\ref{fig_haderiv_hsi}(a) since
no source at this energy band was detectable at that time.

These energy bands were chosen for comparison with $\mathrm{d}I_{\mathrm{H}\alpha}/\mathrm{d}t$
for two reasons. First, they
were above the energies where the thermal component dominates X-ray emission, i.e. $\gtrsim12\,\mbox{keV}$.
Secondly, the electrons bombarding the
chromosphere need to have an energy greater than 25~keV
to penetrate into the layers of H$\alpha$
formation, e.g. in the F1 flare model by Machado et al. (1980).
The RHESSI CLEAN images were reconstructed from data accumulated in 1~spin period.
The centre of each RHESSI time interval corresponds to the observational times of
H$\alpha$ and $\mathrm{d}I_{\mathrm{H}\alpha}/\mathrm{d}t$.

The sites of  the fast H$\alpha$ intensity changes located in the northern part of the flare
lie on the boundaries or inside of X-ray sources in the
25~-~40~keV and/or 40~-~70~keV energy range (see Figures~\ref{fig_haderiv_hsi}(b) and (c)).
However, we have found other areas
that reveal abrupt changes in H$\alpha$ intensity but are located
well away from X-ray emission (e.g. Figure~\ref{fig_haderiv_hsi}(a)).
Some of these areas even precede the position of
X-ray emission as it moves to the south in time (Figures~\ref{fig_haderiv_hsi}(c) and (d)).
The difference in the locations of $\mathrm{d}I_{\mathrm{H}\alpha}/\mathrm{d}t$ and the X-ray sources
cannot be explained by a spatial correlation with the X-ray sources at lower or higher energies.
There is no significant shift of the X-ray sources in the energy ranges from 3~-~7~keV
to 7~-~12~keV or from 40~-~70~keV to higher energy bands; the sources at energies from 7 to 25~keV
are not located closer to the $\mathrm{d}I_{\mathrm{H}\alpha}/\mathrm{d}t$
(see Figures~\ref{fig_ha_hsi_mdi_low} and \ref{fig_ha_hsi_mdi_high}) either.

The fact that at 08:25:20~UT there are
no locations of $\mathrm{d}I_{\mathrm{H}\alpha}/\mathrm{d}t > 0.03\,\mathrm{I}_\mathrm{q}\,\mathrm{s}^{-1}$
along the whole flare area may be the result of the saturation of the H$\alpha$ data at this time
(areas covered by horizontal lines in Figure~\ref{fig_haderiv_hsi}). 
Although the described analysis technique does not eliminate the saturated pixels, they only affect 
the adjacent pixels and cause a decrease in $\mathrm{d}I_{\mathrm{H}\alpha}/\mathrm{d}t$.
\section{Discussion}\label{discussion}
The photon spectrum at the burst maximum of the August 20, 2002 flare is,
to our knowledge, the flattest ($\gamma_1=1.8$) yet detected by RHESSI with its high
energy resolution.  To eliminate any distortion of the spectra, the background subtraction also included 
the estimation of the electron precipitation component, which has been done for the RHESSI 
data for the first time.
The solar origin of such an unusually flat spectrum was confirmed by a comparison 
with the spectrum constructed from the RHESSI PIXON images.

Using a new method of the photospheric albedo correction independent 
of the spectral form of the primary spectrum,
we have proved the importance of  the photospheric albedo correction in the case
of the analysed flare.
We have shown that in this case of flat X-ray spectra, the photospheric albedo significantly
affects the determination of the mean electron flux distribution 
$\bar{n} V \bar{F}(E)$, mainly the value of the low-energy cutoff.
The contamination of the flare spectra by the photospheric albedo photons may result in dips 
in $\bar{n} V \bar{F}(E)$
indicated both by the forward fitting and inversion methods.
Such dips could be misinterpreted as spectral features inherent to the acceleration/propagation processes of
the energetic particles in the solar flares.

Flares of flat hard X-ray spectra ($\gamma\leq 2$) were previously analysed
e.g. by Nitta, Dennis, and Kiplinger (1990) and F\'arn\'ik, Hudson, and Watanabe (1997).
All of these flares have weak thermal components (GOES class M3 or lower) but intense non-thermal
components up to several hundred keV with a flat part extending
down to $\sim 10$~keV.
As was shown here, such photon spectra may correspond to a flat electron flux
distribution with the low-energy cutoff close to the energy of the thermal electrons.

For the flares with spectral index around 3 it is the 
low-energy cutoff which determines the total energy content in the flare. This
happens when the largest contribution comes from the low energy particles. 
However, for the flares with flat spectra $\delta_1\lesssim 1$ it is not the case.
Assuming the collisional thick-target scenario (Brown 1971), one can estimate
the energy flux of the non-thermal electrons injected into the plasma.
In this model, the injected electron flux distribution 
$\mathcal{F}(E_0)\,\mbox{[electrons s}^{-1}\,\mbox{kev}^{-1}\mbox{]}$ is related to $\bar{F}(E)$ of 
Equation~(\ref{ffthin}) as (Brown and Emslie, 1988)
\begin{equation}
\mathcal{F}(E_0)  \sim \bar{F}(E_0) E_0^{-2} 
\left(\mbox{d}\ln\bar{F}(E)/\mbox{d}\ln E + 1\right)_{E=E_0}\sim \bar{F}(E_0)E_0^{-2}\,.
\end{equation}
Relative change of the non-thermal energy flux $\mathfrak{F}_{E_\mathrm{low}}\,\mbox{[keV s}^{-1}\,\mbox{]}$ can
be expressed 
\begin{equation}
\mathfrak{F}_{E'_\mathrm{low}} / \mathfrak{F}_{E_\mathrm{low}}\sim (E'_\mathrm{low}/E_\mathrm{low})^{-\delta_1}\qquad
\mathfrak{F}_{E_\mathrm{low}}=\int_{E_\mathrm{low}}^{E_\mathrm{high}}\mathcal{F}(E_0)E_0\,\mbox{d}E_0\,.
\end{equation}
Adopting the spectral index $\delta_1=0.4$ and the low-energy cutoffs, $E'_\mathrm{low}=10$~keV, $E_\mathrm{low}=30$~keV,
consistent with the inferred flat spectrum,
the ratio is $\sim 2$. Thus, due to the flatness of $\bar{n} V \bar{F}(E)$,
the low values of the low-energy cutoff do not cause a significant increase in the non-thermal energy flux.

The flat distribution function of non-thermal electrons,
especially at low energies, implies that its derivative in the
energy as well as in the velocity space is close to zero or even
positive. Such a distribution function can be formed directly in
the acceleration region (e.g. by the acceleration in the direct
electric field -- Holman, 1985) and/or modified by propagation effect, particle
collisions in dense plasma, and by interactions of accelerated
electrons with plasma waves (Melnik and Kontar, 2003).
If the plasma waves are generated, then the radio emission can be
observed at high frequencies, as in our case. However, detailed
conditions in the radio source needed to confirm this idea are not
known to us.

Our analysis of the characteristics of the hard X-ray sources and the H$\alpha$ and radio emission
revealed several facts about the flare process. First,
the  presented flare was weak in the metric range
during the whole event, except for the short-lasting relativistic type III burst
(radio manifestation of relativistic electrons; Klassen, Karlick{\' y}, and Mann, 2003).
It resembles the March~6, 1989 $\gamma$-ray flare, which started
as silent in radio waves below 800~MHz. Similarly to Rieger, Treumann, and Karlick\'{y} (1999),
we explain the lack of metric radio emission by closed geometry of the magnetic structure 
and a low altitude of the flare process.
Additionally, the high value of the turn-over frequency, 
19.6 - 50~GHz,
of the radio emission
indicates a strong magnetic field or a large depth of the emitting source (Benka and Holman, 1992), 
thus supporting this conclusion. Furthermore, the coalignment of the H$\alpha$ and hard X-ray 
images with the magnetic field extrapolation shows that both the H$\alpha$ kernels 
and the hard X-ray sources are located at the footpoints of the low altitude 
magnetic loops which are the part of the quadrupolar structure (Figure~\ref{fig_mag}).

Second, our results concerning the time and spatial correlation of the hard X-ray 
sources and H$\alpha$ emission are partly in agreement with the prediction of solar flare models.
We have found that most of the X-ray emission maxima, mainly those at the energies above 12~keV 
(corresponding to the X-ray emission of the non-thermal electrons),
tend to be located on the northern and eastern edges of the H$\alpha$
emission (see Figures~\ref{fig_ha_hsi_mdi_low} and \ref{fig_ha_hsi_mdi_high}).
This is consistent with the expectation that X-ray sources should be located
on the external edges  of H$\alpha$ ribbons and similar to those reported by Czaykowska et al. (1999).

Flare models also suggest that fast changes of  H$\alpha$
intensities might be located at positions of X-ray sources, i.e. at the areas
which are bombarded by the suprathermal electrons. In some cases,
this suggestion is confirmed by our analysis.
However, at some times the positions of
the fast changes of  H$\alpha$ intensities do not spatially coincide with the X-ray sources
(see southern locations of
$\mathrm{d}I_{\mathrm{H}\alpha}/\mathrm{d}t$ in the bottom row of
Figure~\ref{fig_haderiv_hsi}).
We propose that such H$\alpha$ brightenings might be caused either by
thermal conduction from a hot plasma located above, or by soft X-ray irradiation.
Using the lowest energy ranges of RHESSI (3~-~7~keV),
we tried to find such a hot plasma at these places, but without any success.
However, we must point out that the time resolution of $\mathrm{d}I_{\mathrm{H}\alpha}/\mathrm{d}t$
was 9~s. Therefore to confirm the existence of these areas,
more analyses and better quality data are needed.

Finally, the time evolution of the positions of the hard X-ray sources may indicate
the movement of the reconnection site. This can be seen e.g. at 08:25:14~UT 
(Figures~\ref{fig_ha_hsi_mdi_low} and \ref{fig_ha_hsi_mdi_high}) when
the hard X-ray sources of different energies are distributed
along the H$\alpha$ ribbon in such a way that the sources of higher energy  are located further to the south-east
direction. Such difference in the positions could be expected in a single flare loop
when it would correspond to the height distribution of the hard X-ray sources along the loop
as found in Aschwanden, Brown, and Kontar (2002). However, this is not
the case here because the extrapolation of the magnetic field shows no magnetic loop along the H$\alpha$ ribbon.
The displacement of the H$\alpha$ ribbon and X-ray sources from the quadrupolar configuration is not due to
the positional errors in the H$\alpha$, MDI, and RHESSI images, which are a factor of 2 
and 4 smaller, respectively.
The offset could be caused by neglecting the electric currents in the current-free extrapolation.
Although the linear force-free extrapolations show similar results, we cannot 
exclude the presence of localised currents which deform the magnetic field structure.
\section{Conclusions}
We have demonstrated that the influence of the photospheric albedo on solar flare X-ray 
spectra must be considered in order to correctly assess the models of electron acceleration 
and propagation and generation of hard X-ray emission. The inferred flat mean electron flux 
distribution could be the result of the acceleration process
and/or modified by particle-wave interactions. This explanation, however, remains speculative.
The accumulated information from the multi-wavelength observations of this flare shows
that the flare process took place in the low altitude magnetic loops.
Our analysis of H$\alpha$ emission may indicate that the response of the lower atmosphere 
to the flare energy release is not restricted to the sites of propagation of the accelerated electrons.
\begin{acknowledgements}
One of the authors, JK, would like to give special
thanks to the RHESSI team at GSFC for their help and support during
her stay there. We are thankful to  W.~Otruba (Kanzelh\"{o}he Solar
Observatory, Austria) and A.~Veronig for providing H$\alpha$ data and
for valuable discussion on their analysis. We also thank T.~Metcalf
for help with the PIXON reconstruction, D.~Zarro for preprocessing
TRACE data, and P.~D\'{e}moulin for providing his code for the
magnetic field extrapolation. Furthermore, we thank P.~Messmer
and T. L\"{u}thi for the Z\"{u}rich and Bern radio data, respectively.

This work  was supported  by NATO Science Fellowships Programme, the contract
NAS5 - 01160,
the grants  IAA3003202 and IAA3003203 of the  Academy of
Sciences of the  Czech Republic, the grants 205/02/0980 and 205/04/0358
of the Grant Agency of the  Czech Republic, and the 
projects K2043105 and AV0Z10030501 of the Astronomical Institute.
\end{acknowledgements}

\end{article}

\begin{thebibliography}{}
\bibitem[]{} Aschwanden, M.~J., Brown, J.~C., and Kontar, E.~P.: 2002, {\it Solar Phys.}, {\bf 210}, 383.
\bibitem[]{} Alexander, D. and Metcalf, T.~R.: 1997, {\it ApJ}, {\bf 489}, 422.
\bibitem[]{} Alexander, R.~C. and Brown, J.~C.: 2002, {\it Solar Phys.}, {\bf 210}, 407.
\bibitem[]{} Asai, A., Masuda, S., Yokoyama, T., Shimojo, M., Isobe, H.,
Kurokawa, H., and Shibata, K.: 2002, {\it ApJL}, {\bf 578}, L91
\bibitem[]{} Bai, T. and Ramaty, R.: 1978, {\it ApJ}, {\bf 219}, 705.
\bibitem[]{} Benka, S.~G. and Holman, G.~D.: 1992, {\it ApJ}, {\bf 391}, 854.
\bibitem[]{} Brown, J.~C.: 1971, {\it Solar Phys.}, {\bf 18}, 489.
\bibitem[]{} Brown, J.~C. and Emslie, A.~G.: 1988, {\it ApJ}, {\bf 331}, 554.
\bibitem[]{} Brown, J.~C., Emslie, A.~G. and Kontar, E.~P.: 2003, {\it ApJ}, {\bf 595}, L115
\bibitem[]{} Cargill, P.~J. and Priest, E.~R.: 1983, {\it ApJ}, {\bf 266}, 383
\bibitem[]{} Canfield, R.~C. and Gayley, K.~G.: 1987, {\it ApJ}, 363, 1999
\bibitem[]{} Czaykowska,  A., de Pontieu, B., Alexander,  D., and Rank, G.:
1999, {\it ApJ}, {\bf 521}, L75.
\bibitem[] {} D\'{e}moulin, P., Bagal\'{a}, L.~G., Mandrini, C.~H., H\'{e}noux, J.~C.,
and Rovira, M.~G.:  1997, {\it Astron. Astrophys.}, {\bf 325}, 305.
\bibitem[]{} Dennis, B.~R.: 1985, {\it Solar Phys.}, {\bf 100}, 465.
\bibitem[]{} Dennis, B.~R. and Schwartz,  R.~A.: 1989, {\it Solar Phys.}, {\bf 121}, 75.
\bibitem[]{} F\'arn\'ik, F., Hudson, H., and Watanabe, T: 1997, {\it Astron. Astrophys.}, {\bf 320}, 620
\bibitem[]{} Fletcher, L. and Hudson, H.: 2002, {\it Solar Phys.}, {\bf 210}, 307.
\bibitem[]{} Haug, E.: 1997, {\it Astron. Astrophys.}, {\bf 326}, 417.
\bibitem[]{} Heinzel, P.: 1991, {\it Solar Phys.}, {\bf 135}, 65.
\bibitem[]{} Heyvaerts, J., Priest, E.~R., and Rust, D.~M.: 1977, {\it ApJ}, {\bf 216}, 123.
\bibitem[]{} Hirose, S. and  Uchida, Y.: 2002, in P.~C.~H.~Martens and D.~P.~Cauffman (eds), {\it Multi-wavelenght
observations of coronal structure  and dynamics}, Yohkoh 10th Anniversary Meeting, p. 181.
\bibitem[]{} Holman, G.~D.: 1985, {\it ApJ}, {\bf 293}, 584.
\bibitem[]{} Holman, G.~D., Sui, L., Schwartz, R.~A., and Emslie, A.~G.: 2003, {\it ApJ}, {\bf 595}, L97.
\bibitem[]{} Hurford, G.~J., Schmahl, E.~J., Schwartz, R.~A., Conway, A.~J. et al.: 2002,
{\it Solar Phys.}, {\bf 210}, 61.
\bibitem[]{} Johns, Ch.~M. and Lin, R.~P.: 1992, {\it Solar Phys.}, {\bf 137}, 121.
\bibitem[]{} Klassen, A., Karlick\'{y}, M., and Mann, G.: 2003, {\it  Astron. Astrophys.}, {\bf 410}, 307.
\bibitem[]{} Karlick\'y, M.: 1997, {\it Space Science Rev.}, {\bf 81}, 143.
\bibitem[]{} Kontar, E.~P., Piana, M., Massone, A.~M., Emslie, A.~G., and
Brown, J.~C.: 2004, {\it Solar Phys.}, {\bf 225}, 293.
\bibitem[]{} Kontar, E.~P., Emslie, A.~G., Piana, M., Massone, A.~M., and
Brown, J.~C.: 2005a, {\it Solar Phys.}, {\bf 226}. 317.
\bibitem[]{} Kontar, E.~P., MacKinnon, A.~L., Schwartz, R.~A., and Brown, J.~C.: 2005b, {\it ApJ}, submitted.
\bibitem[]{} Lampton, M., Margon, B., and Bowyer, S.: 1976, {\it ApJ}, {\bf 208}, 177.
\bibitem[]{} Lin, R.~P. and Schwartz, R.~A.: 1987, {\it ApJ}, {\bf 312}, 462.
\bibitem[]{} Lin, R.~P. et al.:, 2002, {\it Solar Phys.}, {\bf 210}, 3.
\bibitem[]{} Machado, M.~E., Avrett, E.~H., Vernazza, J.~E., and Moyes, R.~W.:
1980, {\it Astron. Astrophys.}, {\bf 242}, 336.
\bibitem[]{} Melnik, V.N. and Kontar, E.P.: 2003, {\it Solar Phys.}, {\bf 215},
335.
\bibitem[]{} Metcalf, T.`R., Hudson, H.~S., Kosugi, T., Puetter, R.~C.,
and Pi\~na, R.~K.: 1996, {\it ApJ}, {\bf 466}, 585.
\bibitem[]{} Nitta, N., Dennis, B.~R., and Kiplinger, A.~L.: 1990, {\it ApJ}, {\bf 353}, 313
\bibitem[]{} Piana, M., Massone, A.~M., Kontar, E.~P., Emslie, A.~G., Brown, J.~C., and Schwartz, R.~A.: 2003,
{\it ApJ}, {\bf 595}, L127.
\bibitem[]{} Poquerusse, M.:  1994, {\it Astron. Astrophys.} {\bf 286},
611.
\bibitem[]{} Press, W.~H., Teukolsky, S.~A., Vetterling, W.~T., and Flannery, B.~P.:
1992, {\it Numerical recepies in C: the art of scientific computing}, 2nd ed.,
Cambridge University Press, p. 684
\bibitem[]{} Puetter, R.~C.~ and Pi\~{n}a, R.~K.: 1994, {\it Experimental Astronomy},
{\bf 3}, 293.
\bibitem[]{} Rieger, E., Treumann, R.~A., and Karlick\'y, M.: 1999, {\it
Solar Phys.}, {\bf 187}, 59.
\bibitem[]{} Schwartz,  R.~A., Csillaghy,  A., Tolbert,  A.~K., Hurford,
G.~J., McTiernan, J., and Zarro,  D.: 2002, {\it Solar Phys.}, {\bf 210},
165.
\bibitem[]{} Smith, D.~M. et al.: 2002, {\it Solar Phys.}, {\bf 210}, 33.
\bibitem[]{} Trottet, G., Rolli, E., Magun, A., Barat, C., Kuznetsov, A., Sunyaev, R., and Terekhov, O.:
2000, {\it Astron. Astrophys.}, {\bf 356}, 1067.
\bibitem[]{} Uchida, Y.: 1996, {\it Adv. Space Res.} {\bf 17}, 19.
\bibitem[]{} Veronig, A., Steinegger, M., Otruba, W., Hanslmeier, A., Messerotti, M.,
       Temmer, M., Gonzi, S., and Brunner, G.: 2000, {\it Hvar. Obs. Bull.}, {\bf 24}, No. 1, 195.
\bibitem[]{} Wang, H., Qiu, J., Denker, C., Spirock, T., Chen, H., and Goode, P.~R.:
      2000, {\it ApJ}, {\bf 542}, 1080.
\end{thebibliography}
\end{document}